\documentstyle{article}
\date{}             
\begin{document}
\title{The strong $\rho NN$ coupling derived from QCD
\thanks{zhusl@itp.ac.cn} }
\author{Shi-Lin Zhu\\
Institute of Theoretical Physics, Academia Sinica\\ 
P.O.BOX 2735, Beijing 100080, P.R.China}
\maketitle
\begin{center}
\begin{minipage}{120mm}
\vskip 0.6in
\begin{center}{\bf Abstract}\end{center}
{\large
We study the two point correlation function of two nucleon currents 
sandwiched between the vacuum and the rho meson state. 
The light cone QCD sum rules are derived for the $\rho NN$ vector and tensor couplings 
simultaneously. The contribution from the excited states and the 
continuum is subtracted cleanly through the double Borel transform 
with respect to the two external momenta, $p_1^2$, $p_2^2=(p-q)^2$.
Our results are: $g_\rho =2.5\pm 0.2$, $\kappa_\rho =(8.0\pm 2.0)$, 
in good agreement with the values used in the nuclear forces.

PACS Indices: 21.30.+y; 13.75.Cs; 12.38.Lg

}
\end{minipage}
\end{center}

\large
\section{Introduction}
\label{sec1}
Quantum Chromodynamics (QCD) is asymptotically free and its high energy 
behavior has been tested to two-loop accuracy. On the other hand, the 
low-energy behavior has become a very active research field in the 
past years. Various hadronic resonances act as suitable labs for 
exploring the nonperturbative QCD. Among which, 
the inner structure of nucleon and mesons and their interactions
is of central importance in nuclear and particle physics. 

Internationally there are a number of experimental  
collaborations, like TJNAL (former CEBAF), 
COSY, ELSA (Bonn), MAMI (Mainz) and Spring8 (Japan), 
focusing on the nonperturbative QCD dynamics. Especially
the Mainz research project MAMI (Mainzer Mikrotron) with its planned extension
from a 855 MeV to 1.5 GeV electron c.w. accelerator 
and the Japan Hadron Facility (JHF) at Spring8 
will extensively study the photo- and electro-production of vector mesons
off nucleons. 

Moreover, the strong $\rho NN$ couplings are , like $\pi NN$ and 
$\pi N\Delta$ couplings, the basic inputs for the description of nuclear 
forces in terms of meson exchange between nucleons. So far the linkage 
between the underlying theory QCD and the phenomenological $\rho NN$ 
couplings has not been made. Especially the commonly adopted 
tensor-vector ratio $\kappa_\rho =6.8$ is much larger than the vector meson 
dominance model (VDM) result $\kappa_v =3.7$. One wonders whether it is
feasible to calculate the $\rho NN$ couplings directly 
with the fundamental theory QCD?

Although it is widely accepted that QCD is the underlying theory of 
the strong interaction, the self-interaction of the gluons causes 
the infrared behavior and the vacuum of QCD highly nontrivial. 
In the typical hadronic scale QCD is nonperturbative which makes 
the first principle calculation of these couplings unrealistic except the
lattice QCD approach, which is very computer time consuming. 
Therefore, a quantitative calculation of the $\rho NN$
couplings with a tractable and reliable theoretical approach proves valuable.

The method of QCD sum rules (QSR), as proposed
originally by Shifman, Vainshtein, and Zakharov \cite{SVZ} and adopted,
or extended, by many others \cite{RRY,IOFFE,BALIT}, are very 
useful in extracting the low-lying hadron masses and couplings.
In the QCD sum rule approach the nonperturbative QCD effects 
are taken into account through various condensates 
in the nontrivial QCD vacuum. A recent review of QSR is given 
by Shifman \cite{shifman}. In this work we shall use the light cone 
QCD sum rules (LCQSR) to calculate the $\rho NN$ couplings.

The LCQSR is quite different from the conventional QSR, which is 
based on the short-distance operator product expansion. 
The LCQSR is based on the OPE on the light cone, 
which is the expansion over the twists of the operators. The main contribution
comes from the lowest twist operator. Matrix elements of nonlocal operators 
sandwiched between a hadronic state and the vacuum defines the hadron wave
functions. When the LCQSR is used to calculate the coupling constant, the 
double Borel transformation is always invoked so that the excited states and 
the continuum contribution can be treated quite nicely. Moreover, the final 
sum rule depends only on the value of the hadron wave function at a 
specific point, which is much better known than the whole wave function \cite{bely95}. 
In the present case our sum rules involve with the rho
wave function (RWF) $\varphi_{\rho}(u_0 ={1\over 2})$. 
Reviews of the method of LCQSR can be found in \cite{lcsr-1,lcsr-2}.

The LCQSR has been widely used to treat the couplings of pions with hadrons.
Recently the couplings of pions with heavy mesons 
in full QCD \cite{bely95}, in the limit of $m_Q\to \infty$ 
\cite{zhu1}, $1/m_Q$ corrections and mixing effects \cite{zhu3}, the couplings 
of pions with heavy baryons \cite{zhu2}, the $\pi NN$ and $\pi N N^\ast (1535)$
couplings \cite{zhu5}, the $\rho\to\pi\pi$ and $K^\ast\to K \pi$ decays 
\cite{zhu6}, and various semileptonic decays of heavy 
mesons \cite{bagan98} are discussed.

The QCD sum rules were used to analyze the exclusive radiative 
$B$-decays with the help of the light-cone vector meson wave function 
in \cite{braun94}. With the same formalism the off-shell $g_{B^* B\rho}$ 
and $g_{D^* D\rho}$ couplings in \cite{aliev96} and the $\rho$ decay 
widths of excited heavy mesons \cite{zhu4} were calculated.

Our paper is organized as follows.
Section \ref{sec1} is an introduction.
We introduce the two point function for the $\rho NN$
vertex and saturate it with nucleon intermediate states in section \ref{sec2}. 
The definitions of the RWFs and the formalism of LCQSR 
are presented in section \ref{sec3}.
In the next section we present the LCQSR for the $\rho NN$ coupling. 
In section \ref{sec5} we present some discussions of these RWFs and 
their values at the point $u_0 ={1\over 2}$.
We make the numerical analysis and a short discussion in section \ref{sec6}. 

\section{Two Point Correlation Function for the $\rho NN$ coupling}
\label{sec2}

Many authors have studied the strong $\rho NN$ couplings.
It was pointed out that the inclusion of an effective $\rho$-pole contribution 
leads to a large value for the tensor-vector coupling ratio 
$\kappa_\rho =6.6\pm 1$ \cite{holer,mergell}, in the dispersion-theoretical 
analysis of the nucleon electromagnetic form factors. The above value is consistent 
with other determination \cite{grein,furuichi}. Brown and Machleidt 
have discussed the evidence for a strong $\rho NN$ coupling from the 
measurement of $\epsilon_1$ parameter in NN scattering \cite{machleidt}.
Brown, Rho and Weise suggested that $\kappa_\rho =2\kappa_v$ is consistent 
with a quark core radius of $0.5$fm for the nucleon and an equal factorization
of the baryon charge between the quark and meson cloud in the two-phase 
Skyrme model \cite{brown-rho}. Recently Wen and Hwang used the external 
field method in QCD sum rules to study the $\rho NN$ 
couplings \cite{hwang}. They obtained $\kappa_\rho =3.6$, 
in agreement with VDM result $\kappa_v =3.7$ 
and $\kappa_s =-0.12$. In their work the authors introduced the 
vector-like $\rho$-quark 
interaction Lagrangian by hand and treated the vector meson quark coupling
as free parameter. In other words, the $\rho NN$  
vector and tensor couplings can not be determined simultaneously.

We shall calculate the $\rho NN$ vector and tensor couplings
simultaneously using vector meson light cone wave functions up to twist four, 
which will result in a reliable extraction of $\kappa_\rho$.

We start with the two point function 
\begin{equation}
\Pi (p_1,p_2,q) = \int d^4 x e^{ip x} 
\left \langle 0 \vert {\cal T}
\eta_p (x)  {\bar{\eta_p}} (0) \vert \rho^0 (q) \right \rangle
\label{three-point}
\end{equation}
with $p_1 =p$, $p_2 = p-q$ and 
the Ioffe nucleon interpolating field \cite{IOFFE}
\begin{equation}\label{cur1}
\eta_p (x) = \epsilon_{abc}  [  u^a (x) {\cal C} \gamma_\mu
u^b (x) ] \gamma_5 \gamma^\mu d^c (x) \; ,
\label{eq4}
\end{equation}
\begin{equation}
{\bar\eta}_p(y) = \epsilon_{abc}[{\bar u}^b(y) \gamma_\nu C 
         {\bar u}^{aT}(y) ] {\bar d}^c(y) \gamma^\nu \gamma^5\; ,
\label{eq5}
\end{equation}
where $a,b,c$ is the color indices
and ${\cal C} = i \gamma_2 \gamma_0$ is the charge conjugation matrix.
For the neutron interpolating field, $u \leftrightarrow d$.

The rho nucleon couplings are defined by the $\rho NN$ interaction:
\begin{equation}
{\cal L}_{\rho NN} =  g_\rho \rho^\mu {\bar N}
[ \gamma_\mu +\kappa_\rho {i\sigma_{\mu\nu} q^\mu\over 2m_N} ]N
\; .
\label{eq5a}
\end{equation}
$\rho_\mu$ is an isovetor in (\ref{eq5a}). $g_\rho$ is the rho-nucleon 
vector coupling constant and $\kappa_\rho$ is the tensor-vector ratio.

At the phenomenological level the eq.(\ref{three-point}) can be 
expressed as:
\begin{equation}\label{pole}
\Pi (p_1,p_2,q)  =   i \lambda_N^2  g_\rho e^\mu (\lambda)
{ ({\hat p_1}+m_N)(\gamma_\mu  +\kappa_\rho {i\sigma_{\mu\nu} q^\mu\over 2m_N})
({\hat p_2}+m_N) 
\over{ (p_1 ^2 - m_N ^2) (p_2 ^2 - m_N ^2) }} +\cdots
\end{equation}
where $e^\mu (\lambda)$ is the rho meson polarization vector. 
The ellipse denotes the continuum and the single pole excited states to nucleon
transition contribution. $\lambda_N$ is the overlapping amplitude of 
the interpolating current $\eta_N (x)$ with the nucleon state
\begin{equation}
\left \langle 0 \vert \eta (0) \vert N (p) \right \rangle 
= \lambda_N  u_N (p)
\end{equation}

Expanding (\ref{pole}) with the independent variables $P={p_1 +p_2\over 2}, q$ and
decomposing it into the chiral odd and chiral even part, we arrive at:
\begin{equation}\label{s}
\Pi =\Pi_o +\Pi_e \; ,
\end{equation}
where 
\begin{eqnarray}\label{odd}\nonumber
\Pi_o (p_1,p_2,q)  =  { i \lambda_N^2  g_\rho 
\over{ (p_1 ^2 - m_N^2) (p_2 ^2 - m_N^2) }}  \{
(e\cdot P){\hat P} +{1+\kappa_\rho\over 2}q^2 {\hat e}
&\\ 
-{1+2\kappa_\rho\over 4}(e\cdot q){\hat q}
+i(1+\kappa_\rho )\epsilon_{\mu\alpha\beta\sigma}
e^\mu P^\alpha q^\beta \gamma^\sigma \gamma_5
\}+ \cdots \; , &
\end{eqnarray}
and 
\begin{eqnarray}\label{even}\nonumber
\Pi_e (p_1,p_2,q)  =  { i \lambda_N^2  g_\rho 
\over{ (p_1 ^2 - m_N^2) (p_2 ^2 - m_N^2) }}  \{
(2m_N +{\kappa_\rho\over 2m_N}q^2)(e\cdot P)
&\\ 
-{1+\kappa_\rho\over 2}m_N({\hat e} {\hat q} -{\hat q}{\hat e}) 
-{\kappa_\rho \over 2m_N} (e\cdot P)({\hat q} {\hat p} -{\hat p} {\hat q}) 
\} + \cdots\; , &
\end{eqnarray}
with $q^2 =m_\rho^2$.
We have not kept the single pole terms in (\ref{odd}) and (\ref{even}) 
since they are always eliminated after making double Borel transformation 
in deriving final LCQSRs.

It was well known that the sum rules derived from the chiral odd tensor 
structure yield better results than those from the chiral even ones
in the QSR analysis of the nucleon mass and magnetic moment 
\cite{IOFFE,zsl-mag}. We shall consider chiral odd tensor structures
only below.

\section{The Formalism of LCQSR and Rho Wave Functions}
\label{sec3}
Neglecting the four particle component of the rho wave function, 	
the expression for $\Pi (p^2_1,p^2_2,q^2)$ 
with the tensor structure at the quark level reads,
\begin{eqnarray}\label{quark} \nonumber 
\int e^{ipx} dx
\langle 0| T\eta_p(x) {\bar \eta}_p(0) |\rho^0 (q)\rangle = 
-4\epsilon^{abc}\epsilon^{a^{\prime} b^{\prime} c^{\prime}}
\gamma_5 \gamma_{\mu} iS_d^{a a^{\prime}}(x) 
&\\
\gamma_{\nu} C 
{\langle 0| d^c (x) {\bar u}^{c^{\prime}}(0)|\rho^+ (q)\rangle}^T
C \gamma_{\mu} i{S^T_u}^{b b^{\prime}}(x) \gamma_{\nu}\gamma_5 
&\; ,
\end{eqnarray}
where $iS(x)$ is the full light quark propagator with both perturbative  
term and contribution from vacuum fields. 
\begin{eqnarray}\label{prop}\nonumber
iS(x)=\langle 0 | T [q(x), {\bar q}(0)] |0\rangle 
=i{{\hat x}\over 2\pi^2 x^4} 
-{\langle {\bar q} q\rangle  \over 12}
-{x^2 \over 192}\langle {\bar q}g_s \sigma\cdot G q\rangle &\\ 
+{g_s\over 16\pi^2}\int^1_0 du \{
2 (1-2u) x_\mu\gamma_\nu +i\epsilon_{\mu\nu\rho\sigma}\gamma_5 
\gamma^\rho x^\sigma \} G^{\mu\nu} 
+\cdots  & \; .
\end{eqnarray}
where we have introduced ${\hat x} \equiv x_\mu \gamma^\mu$. 
In our calculation we take the tiny current quark mass to be zero.

The relevant feynman diagrams are presented in FIG 1. The squares 
denote the rho wave function (RWF). The broken solid line, broken curly line
and a broken solid line with a curly line attached in the middle
stands for the quark condensate, gluon condensate 
and quark gluon mixed condensate respectively. 

By the operator expansion on the light-cone
the matrix element of the nonlocal operators between the vacuum and 
rho meson defines the two particle rho wave function.
Up to twist four the Dirac components of this wave function can be 
written as follows \cite{braun94,braun96,braun98}.
For the longitudinally polarized rho mesons,
\begin{eqnarray} \nonumber 
\langle 0|\bar u(0) \gamma_{\mu} 
d(x)|\rho^-(q,\lambda)\rangle = 
 f_{\rho} m_{\rho} \{ q_{\mu}
\frac{e^{(\lambda)}\cdot x}{q \cdot x}
\int_{0}^{1} \!du\, e^{-iu q\cdot x} 
[\phi_{\parallel}(u, \mu^{2}) +\frac{m^2_\rho x^2}{4}  A(u,\mu)]
&\\   
+ (e^{(\lambda)}_{\mu}- q_\mu{e^{(\lambda)}x\over qx})
\int_{0}^{1} \!du\, e^{-iu q\cdot x} g^v (u, \mu^{2}) 
- \frac{1}{2}x_{\mu}
\frac{e^{(\lambda)}\cdot x }{(q \cdot x)^{2}} m_{\rho}^{2}
\int_{0}^{1} \!du\, e^{-iu q\cdot x} C (u, \mu^{2}) \} \; , &
\label{rwf3}
\end{eqnarray}
and 
\begin{equation}
\langle 0|\bar u(0) \gamma_{\mu} \gamma_{5}
d(x)|\rho^-(q,\lambda)\rangle = -\frac{1}{4}f_{\rho} 
m_{\rho} \epsilon_{\mu}^{\phantom{\mu}\nu \alpha \beta}
e^{(\lambda)}_{\perp \nu} q_{\alpha} x_{\beta}
\int_{0}^{1} \!du\, e^{-iu q\cdot x} g^a (u, \mu^{2}) \,
\label{rwf4}
\end{equation}
 where
\begin{equation}
C (u) = g_3(u)+\phi_\parallel(u) -2 g^{(v)}_\perp(u)\; .
\end{equation}

The link operators 
$ \mbox{Pexp}\left[ig\int_0^1 d\alpha\, x^\mu A_\mu(\alpha x)\right]  $
are understood in between the quark fields.
The distribution amplitudes describe the probability amplitudes to 
find the $\rho$ in a state with quark and antiquark carrying
momentum fractions $u$ (quark) and $1-u$ (antiquark), respectively, and
have a small transverse separation of order ${1\over \mu}$ \cite{braun98}.

The vector and tensor decay constants $f_\rho$ and $f_\rho^T$ are defined as
\begin{equation}
\langle 0|\bar u(0) \gamma_{\mu}
d(0)|\rho^-(q,\lambda)\rangle  =  f_{\rho}m_{\rho}
e^{(\lambda)}_{\mu}\; .
\label{frho}
\end{equation}

All distributions $\phi=\{\phi_\parallel, g^v,g^a,A,C\}$ are normalized as
\begin{equation}
\int_0^1\!du\, \phi(u) =1,
\label{norm}
\end{equation}

The twist-three three-particle quark-antiquark-gluon distributions are \cite{braun98}:
\begin{equation} 
\langle 0|\bar u(0) \gamma_\alpha g_sG_{\mu\nu}(ux)
         d(x)|\rho^-(q,\lambda)\rangle  = 
      iq_\alpha[q_\mu e^{(\lambda)}_{\perp\nu}-
      q_\nu e^{(\lambda)}_{\perp\mu}]
      f_{3\rho}^V{\cal V}(u,qz)+\cdots \; ,
\label{rwf-v}
\end{equation}
\begin{equation} 
\langle 0|\bar u(0) \gamma_\alpha \gamma_5 g_s{\tilde G}_{\mu\nu}(ux)
d(x)|\rho^-(q,\lambda)\rangle  = q_\alpha[q_\nu e^{(\lambda)}_{\perp\mu}-
 q_\mu e^{(\lambda)}_{\perp\nu}] f_{3\rho}^A{\cal A}(u,qx)+\cdots \; ,
\label{rwf-a}
\end{equation}
where the operator $\tilde G_{\alpha \beta}$  is the dual of $G_{\alpha \beta}$:
$\tilde G_{\alpha \beta}= {1\over 2} \epsilon_{\alpha \beta \delta \rho} 
G^{\delta \rho} $ and the ellipses denote higher twists contribution.

The following shorthand notation for the integrals defining three-particle 
distribution amplitudes is used:
\begin{equation}
   {\cal F}(u,qx) \equiv  \int {\cal D}\underline{\alpha}
   \,e^{-iq\cdot x(\alpha_3+u\alpha_g)}{\cal F}(\alpha_1,\alpha_g,\alpha_3).
\end{equation}
Here ${\cal F} = \{{\cal V,A}\}$ refers to the vector and
axial-vector distributions, 
$\underline{\alpha}$ is the set of three 
momentum fractions: $\alpha_3$ ($d$ quark), $\alpha_1$ ($u$ quark)
and $\alpha_g$ (gluon), and the integration measure is defined as
\begin{equation}
 \int {\cal D}\underline{\alpha} \equiv \int_0^1 d\alpha_1
  \int_0^1 d\alpha_3\int_0^1 d\alpha_g \,\delta(1-\sum \alpha_i).
\label{eq:measure}
\end{equation}
The normalization constants $f_{3\rho}^V, f_{3\rho}^A, f_{3\rho}^T$ 
are defined in such a way that 
\begin{eqnarray}
 \int\! {\cal D}\underline{\alpha}\, (\alpha_3-\alpha_1)\,{\cal V}
    (\alpha_1,\alpha_g,\alpha_3) &=&1,
\\
\int\! {\cal D}\underline{\alpha}\,{\cal A} (\alpha_1,\alpha_g,\alpha_3) &=&1.
\end{eqnarray} 
The function ${\cal A}$ is symmetric and the functions ${\cal V}$
is antisymmetric under the interchange $\alpha_1 \leftrightarrow \alpha_3$
in the SU(3) limit (\cite{CZreport,braun98}), which follows from the
G-parity transformation property of the corresponding matrix elements.

In the infinite momentum frame the RWFs $ \phi_\parallel$ are
associated with the leading twist two operator, 
$A(u)$ correspond to twist four operators, 
and $g^a (u), g^v(u)$ to twist three ones. 
The three particle RWFs ${\cal V}, {\cal A}$ are of twist three.
Details can be found in \cite{braun98}.

\section{The LCQSR for the $\rho NN$ coupling}
\label{sec4}
Expressing (\ref{quark}) with the longitudinal rho wave functions (LRWFs), we 
can obtain the expressions for the correlator in the coordinate space.
Up to dimension six the gluon condensate is the only relevant condensate
contributing to the chiral odd tensor structure. 
Yet the gluon condensate always appears with a large suppression factor, 
which arises from the two-loop internal momentum integration in the diagram 
(d) in FIG 1. Its contribution is quite small, which is confirmed by our 
detailed calculation. For example, after double Borel transformation, 
diagram (d) is suppress by a factor ${ \langle g_s^2 G^2\rangle \over 24 M_B^4}$,
where $\langle g_s^2 G^2\rangle =0.48$GeV$^4$ and $M_B^2\sim 1.5$GeV$^2$.

Diagram (a) involves with two-particle LRWFs.
After tedious but straightforward calculation, we get:
\begin{eqnarray}\label{long-1}\nonumber
\Pi^2_o (p_1, p_2, q) = \int d^4 x \int_0^1 du e^{i(p-uq) x} f_\rho m_\rho\{
-{ g^a (u)\over 2\pi^4 x^6}  \epsilon_{\alpha\beta\mu\sigma}
e^\beta q^\mu x^\sigma \gamma^\alpha \gamma_5
&\\  \nonumber
-{4\over \pi^4 x^6} \{[\phi_\parallel (u)-g^v (u)]
{\hat q} {(e\cdot x) \over (q\cdot x)} + {\hat e} g^v (u) \}
+{2\over \pi^4 x^6} m_\rho^2 g_3 (u) {(e\cdot x)\over (q\cdot x)^2} {\hat x}
&\\ 
-{1\over \pi^4 x^4} m_\rho^2 A(u) {\hat q} {(e\cdot x) \over (q\cdot x)} 
-{ g^a (u)\over 768\pi^4 x^2}\langle g_s^2 G^2\rangle 
\epsilon_{\alpha\beta\mu\sigma}
e^\beta q^\mu x^\sigma \gamma^\alpha \gamma_5
+\cdots \} \; .
\end{eqnarray}
The LRWFs can be found in the previous section.
Diagram (b) is associated with vacuum gluon fields. But its 
contribution vanishes due to isospin symmetry. Our explicit
calculation confirms it.

We frequently use integration by parts to absorb the factors 
$1/(q\cdot x)$ and $1/(q\cdot x)^2$, which leads to the 
integration of RWFs. For example, 
\begin{equation}\label{integration}
\int_0^1 {e^{-iu q\cdot x}\over  q\cdot x }\psi (u)  du =
i\int  e^{-iu q\cdot x} \Psi (u)  du +
\Psi (u) e^{-iu q\cdot x}|_0^1 \; ,
\end{equation}
where the functions $\Psi(u)$ is defined as:
\begin{equation}
\Psi (u)=+\int_0^{u} \psi(u)du \; .
\end{equation}
Note the second term in (\ref{integration}) vanishes 
after double Borel transformation or due to 
$\phi_\rho (u_0) =\Psi (u_0)=0$ at end points $u_0 =0, 1$.

We first finish Fourier transformation. The formulas are: 
\begin{equation}\label{ft1}
\int {e^{ipx}\over (x^2)^n} d^D x \to i (-1)^{n+1} 
{2^{D-2n}\rho^{D/2} \over (-p^2)^{D/2 -n}} {\Gamma (D/2 -n)\over \Gamma (n)} \; ,
\end{equation}
\begin{equation}\label{ft2}
\int {{\hat x}e^{ipx}\over (x^2)^n} d^D x \to  (-1)^{n+1} 
{2^{D-2n+1}\rho^{D/2} \over (-p^2)^{D/2+1 -n}} {\Gamma (D/2+1 -n)\over \Gamma (n)}
{\hat p} \; ,
\end{equation}
\begin{eqnarray}\label{ft3}\nonumber
\int { x_\mu x_\nu e^{ipx}\over (x^2)^n} d^D x \to 
 i(-1)^n 2^{D-2n+1}\pi^{D/2} \{
{g_{\mu\nu}\over (-p^2)^{D/2+1 -n}} {\Gamma (D/2+1 -n)\over \Gamma (n)}
&\\
+{2p_\mu p_\nu\over (-p^2)^{D/2+2 -n}} {\Gamma (D/2+2 -n)\over \Gamma (n)}
\} & \; .
\end{eqnarray}

The next step is to make double Borel transformation with the variables $p_1^2$ and $p_2^2$
to (\ref{odd}) and (\ref{lo-1})-(\ref{lo-2}).
The single-pole terms in (\ref{pole}) are eliminated. The formula reads:
\begin{equation}\label{double}
{{\cal  B}_1}^{M_1^2}_{p_1^2} {{\cal  B}_2}^{M_2^2}_{p_2^2} 
{\Gamma (n)\over [ m^2 -(1-u)p_1^2-up^2_2]^n }=
(M^2)^{2-n} e^{-{m^2\over M^2}} \delta (u-u_0 ) \; ,
\end{equation}
where $u_0={M^2_1 \over M^2_1 + M^2_2}$, 
$M^2\equiv {M^2_1M^2_2\over M^2_1+M^2_2}$, 
$M^2_1$, $M^2_2$ are the Borel parameters.

Finally we identify the same tensor structures both at the hadronic level 
and the quark gluon level.  
Subtracting the continuum contribution which is modeled by the 
dispersion integral in the region $s_1 , s_2\ge s_0$, we arrive at:
\begin{itemize}
\item $i {\hat e}$
\begin{eqnarray}\label{lo-1}\nonumber
\lambda^2_N \sqrt{2} g_\rho {1+\kappa_\rho\over 2}m_\rho^2
e^{-{ m_N^2\over M^2} } = 
{f_\rho m_\rho\over 2\pi^2} e^{-{u_0(1-u_0)m_\rho^2\over M^2}} \{
g^v (u_0) M^6 f_2 ({s_0\over M^2})
&\\ 
- m_\rho^2 G_3 (u_0) M^4 f_1 ({s_0\over M^2})  
+{g^v (u_0)\over 24}  \langle g_s^2G^2 \rangle
 M^2 f_0 ({s_0\over M^2})
\}&\; ,
\end{eqnarray}
\item $i (e\cdot P)  {\hat P}  $
\begin{eqnarray}\label{lo-2}\nonumber
\lambda^2_N \sqrt{2} g_\rho 
e^{-{ m_N^2\over M^2} } = -e^{-{u_0(1-u_0)m_\rho^2\over M^2}} 
{f_\rho m^3_\rho \over \pi^2} G_3(u_0) M^2 f_0 ({s_0\over M^2})
\; ,
\end{eqnarray}
\end{itemize}
where $q^2=m_\rho^2 $, $f_n(x)=1-e^{-x}\sum\limits_{k=0}^{n}{x^k\over k!}$ 
is the factor used to subtract the continuum, $s_0$ is the continuum threshold.
The sum rules are symmetric with the Borel parameters 
$M_1^2$ and $M_2^2$. It's natural to adopt
$M_1^2=M_2^2=2M^2$, $u_0 ={1\over 2}$.
The functions $G_i (u), i=0, 1, 2, 3, A(u)$ are defined as:
\begin{equation}\label{g0}
G_0 (u)=\int_0^u dt \phi_\parallel (t) \; ,
\end{equation}
\begin{equation}\label{g1}
G_1 (u)=\int_0^u dt g^v (t) \; ,
\end{equation}
\begin{equation}\label{g2}
G_2 (u)=\int_0^u dt A (t) \; ,
\end{equation}
\begin{equation}\label{ggg}
G_3 (u)=\int_0^u dt \int_0^t ds C (s) \; .
\end{equation}

\section{Discussions of RWFs and parameters}
\label{sec5}

The resulting sum rules depend on the RWFs, the integrals and derivatives of them 
at the point $u_0={1\over 2}$. The distribution amplitudes of vector mesons have been 
studied in \cite{CZreport,braun96,braun98} using QCD sum rules. We adopt the 
model RWFs in Refs.\cite{braun100}.

The values of the two-particle RWFs, their derivatives and integrals at the point 
$u_0 ={1\over 2}$ using the form in \cite{braun100} are:
$g^a=1.15 \pm 0.23$, $g^v=0.64$, 
$\phi_\parallel (u_0)=1.1$,
$G_0 (u_0)=0.5$, $G_1 (u_0)=0.5$, $G_2 (u_0)=0.58$, 
$G_3 (u_0)=-0.13$, $A (u_0)=2.18$.

The experimental value for the rho meson mass $m_\rho$ and 
the decay constant $f_\rho$ is $f_\rho =198\pm 7 $MeV
and $m_\rho =770$MeV \cite{review}.

The various parameters which we adopt are 
$s_0=2.25$GeV$^2$, $m_N=0.938$GeV, 
$\lambda_N =0.026$GeV$^3$ \cite{IOFFE} at the scale $\mu =1$GeV.  
The working interval for analyzing the QCD sum rules for nucleons
is $0.9\mbox{GeV}^2 \leq M_B^2\leq 1.8\mbox{GeV}^2$, a standard choice 
for analyzing the various QCD sum rules associated with the nucleon. 

\section{Numerical analysis and results}
\label{sec6}

In order to diminish the uncertainty due to $\lambda_N$, we shall 
divide our sum rules by the famous Ioffe's mass sum rule for the nucleon:
\begin{equation}\label{mass}
32\pi^4 \lambda_N^2 e^{-{ M_N^2\over M^2} }
=M^6 f_2 ({s_0\over M^2})+{b\over 4}M^2 f_0 ({s_0\over M^2})
+{4\over 3}a^2 -{a^2m_0^2\over 3M^2} \; .
\end{equation}

Dividing (\ref{lo-1})-(\ref{lo-3}) by (\ref{mass}), we have 
two new sum rules for $g_\rho (1+\kappa_\rho), g_\rho$. 
The dependence on the continuum threshold $s_0$ and Borel parameter $M^2$
of these sum rules are presented in FIG 2-3. From top to bottom 
the curves correspond to $s_0=2.35, 2.25, 2.15$ respectively. 
These sum rules are stable with reasonable variations 
of $s_0$ and $M^2$ as can be seen in FIG 2-3. 
Numerically we have:
\begin{equation}\label{num-1}
g_\rho (1+\kappa_\rho )= (22 \pm 3)\; ,
\end{equation}
for ${(\ref{lo-1})\over (\ref{mass})}$ and
\begin{equation}\label{num-2}
g_\rho = ( 2.5\pm 0.2)\; ,
\end{equation}
for ${(\ref{lo-2})\over (\ref{mass})}$.

We can also divide (\ref{lo-1}) by (\ref{lo-2}) to get a new sum rule for 
$1+\kappa_\rho$. 
\begin{equation}\label{lo-3}
1+\kappa_\rho =-{ g^v (u_0) M^4 f_2 ({s_0\over M^2})
- m_\rho^2 G_3 (u_0) M^2 f_1 ({s_0\over M^2}) +{g^v (u_0)\over 24}  
\langle g_s^2G^2 \rangle  f_0 ({s_0\over M^2}) \over m_\rho^4
G_3 (u_0)  f_0 ({s_0\over M^2}) } \; .
\end{equation}
The result is presented in FIG 4. Numerically,
\begin{equation}\label{num-3}
1+\kappa_\rho =( 9.0\pm 2.0)\; ,
\end{equation}
which corresponds to
\begin{equation}\label{num-4}
\kappa_\rho =( 8.0\pm 2.0)\; .
\end{equation}

Brown and Machleidt emphasized that the strong $\rho NN$ coupling 
\begin{equation}\label{bm}
{g_\rho^2 (1+\kappa_\rho )^2 \over 4\pi }=(37\pm 13)
\end{equation}
\begin{equation}
\kappa_\rho =(6.6\pm 0.1)
\end{equation}
should be adopted in order to reproduce experimental data \cite{machleidt}.
The vector meson dominance (VMD) model yields 
\begin{equation}\label{vmd}
{g_\rho^2 (1+\kappa_\rho )^2 \over 4\pi }=13.25 \; .
\end{equation}
Our result is 
\begin{equation}\label{zsl}
{g_\rho^2 (1+\kappa_\rho )^2 \over 4\pi }=(39\pm 10) \; ,
\end{equation}
which agrees very well with (\ref{bm}) and deviates strongly from VMD 
prediction (\ref{vmd}).

We have included the uncertainty due to the variation of the continuum 
threshold and the Borel parameter $M^2$ in our analysis. 
Other sources of uncertainty include: (1) the truncation of OPE on the light 
cone and keeping only the few lowest twist operators; (2) the inherent 
uncertainty due to the model RWFs etc. In the present case the major  
uncertainty comes from the RWFs since our final sum rules depends 
both on the value of RWFs and their integrals at $u_0$. 

Our result $\kappa_\rho =( 8.0\pm 2.0)$ is about two times larger than 
that derived in \cite{hwang}, $\kappa_\rho =3.6$ treating the rho meson
as the external field. The reason is two-old. Firstly, the vector-like
$\rho q q$ interaction is assumed in \cite{hwang}. Moreover, for the 
antisymmetric sum rules the susceptibilities $\chi$, $\kappa$ and $\xi$ 
take the same values as in the nucleon magnetic sum rules where 
the electromagnetic field is treated as the external field. 
In our opinion, such a treatment employs the 
vector meson dominance assumption inexplicitly, which may be the 
underlying reason for $\kappa_\rho \approx \kappa_v =3.7$. Secondly, 
the vector meson mass corrections turn out to be large as pointed 
out in \cite{braun100}. This effect is explicitly taken into account 
in our calculation.

In summary we have calculated the $\rho N N$ couplings starting from QCD. 
The continuum and the excited states contribution is subtracted rather 
cleanly through the double Borel transformation in both cases.
Our result strongly supports large value for the tensor-vector 
ratio $\kappa_\rho$ in the nuclear force.

\vspace{0.8cm} {\it Acknowledgments:\/} This work was 
supported by the Postdoctoral Science Foundation of China
and Natural Science Foundation of China.

\vspace{1cm}
{\bf Figure Captions}
\vspace{2ex}
\begin{center}
\begin{minipage}{130mm}
{\sf FIG 1.} \small{
The relevant feynman diagrams for the derivation of the LCQSR for 
$\rho NN$ coupling. The squares denote the rho wave function (RWF). 
The broken solid line, broken curly line
and a broken solid line with a curly line attached in the middle
stands for the quark condensate, gluon condensate 
and quark gluon mixed condensate respectively.  }
\end{minipage}
\end{center}
\begin{center}
\begin{minipage}{130mm}
{\sf FIG 2.} \small{The sum rule for $g_\rho (1+\kappa_\rho )$ 
as a function of the Borel parameter $M^2$ for
${(\ref{lo-1})\over (\ref{mass})}$ with the model RWFs in \cite{braun100}. 
From bottom to top the curves correspond to 
the continuum threshold $s_0 =2.35, 2.25, 2.15$GeV$^2$.}
\end{minipage}
\end{center}
\begin{center}
\begin{minipage}{130mm}
{\sf FIG 3.} \small{ The sum rule for $g_\rho $ 
as a function of $M^2$ and $s_0$ for ${(\ref{lo-2})\over (\ref{mass})}$.
}
\end{minipage}
\end{center}
\begin{center}
\begin{minipage}{130mm}
{\sf FIG 4.} \small{
The sum rule for $1+\kappa_\rho $ 
as a function of $M^2$ and $s_0$ from ${(\ref{lo-1})\over (\ref{lo-2})}$.
}
\end{minipage}
\end{center}

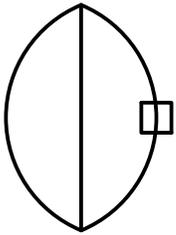
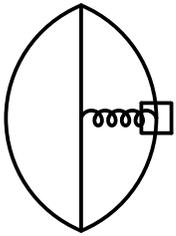
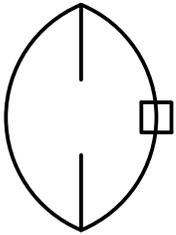
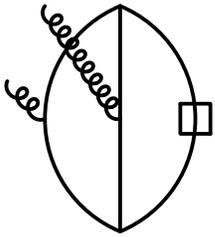
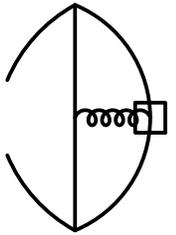
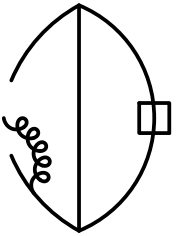
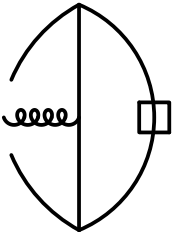
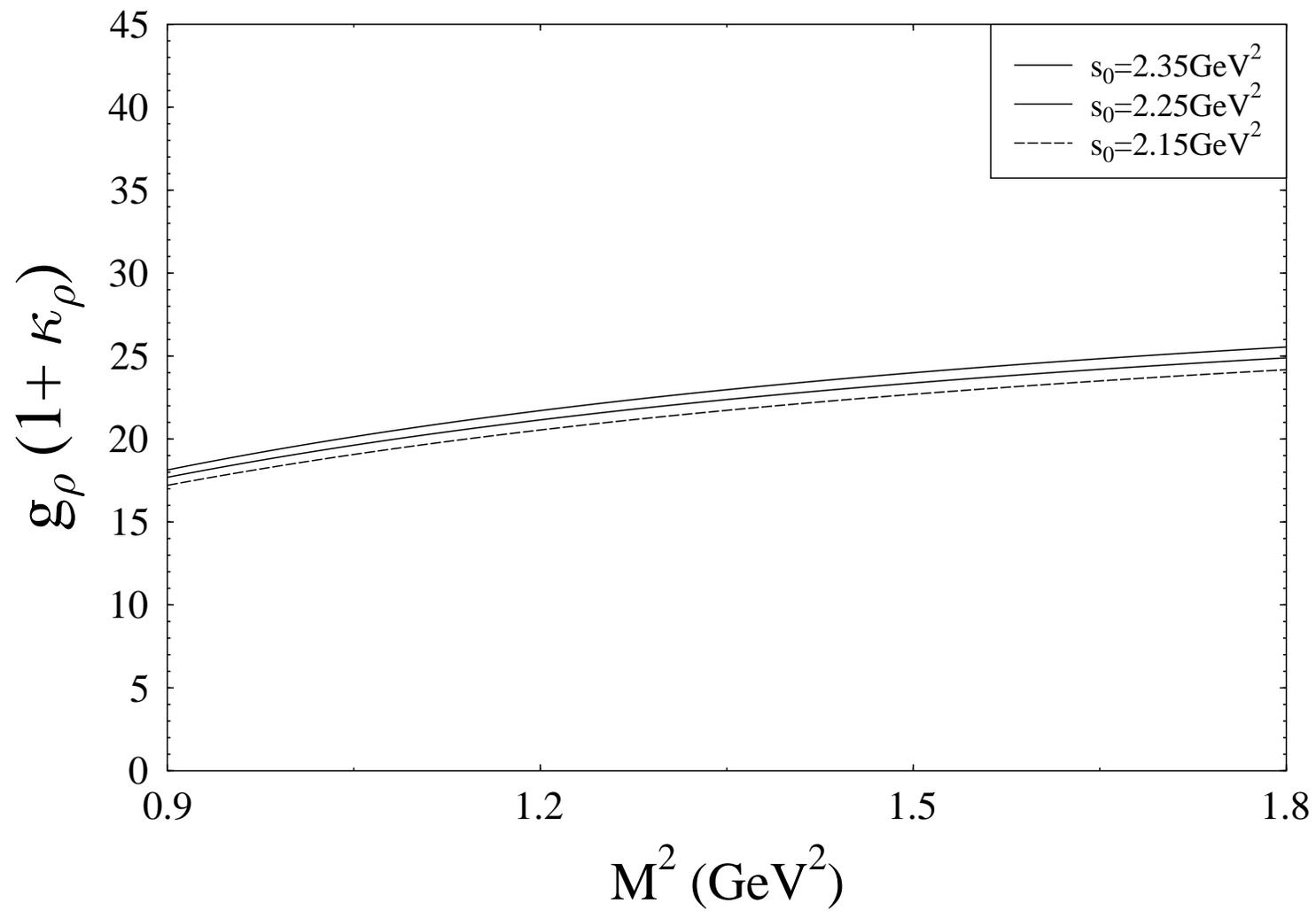
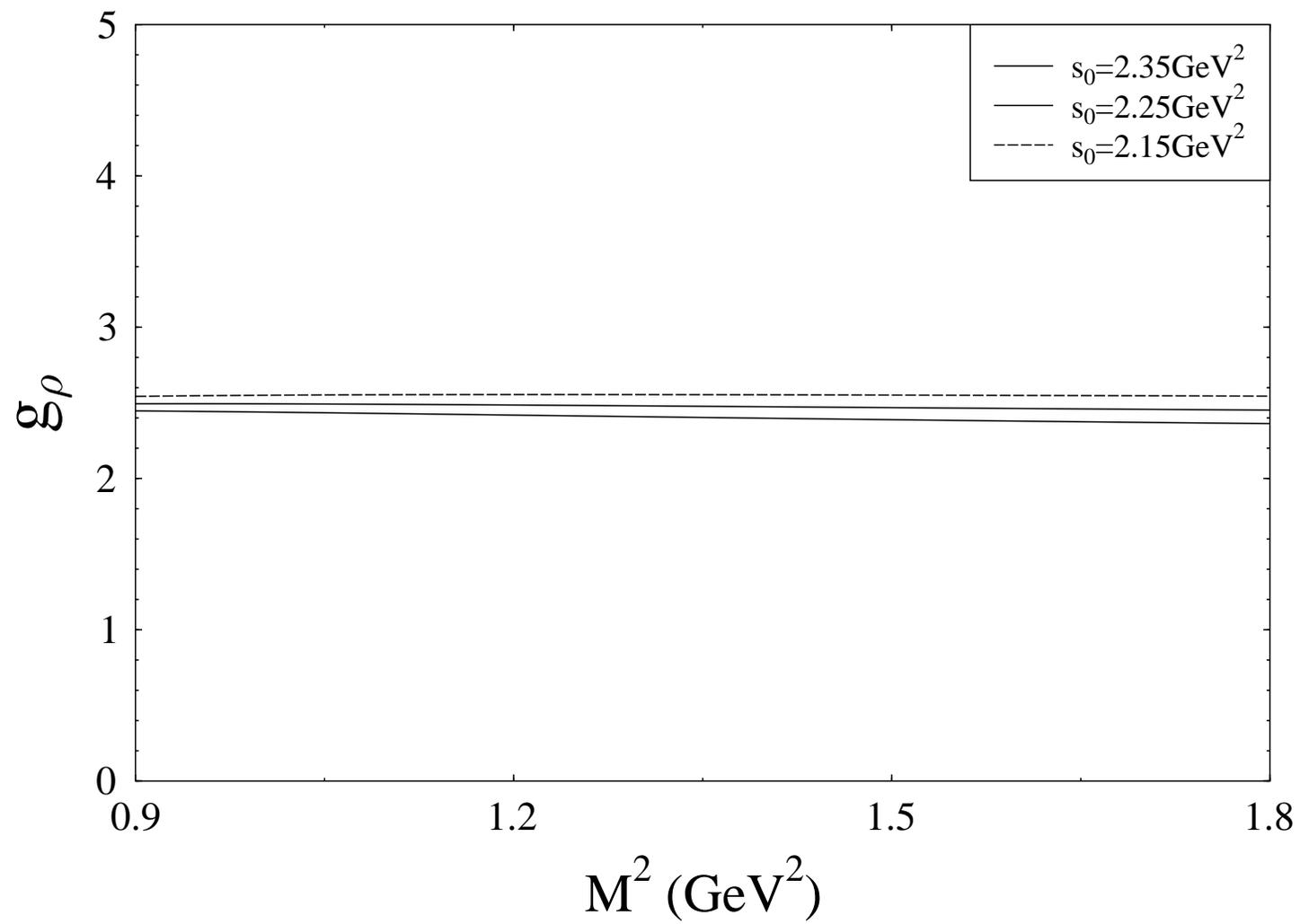
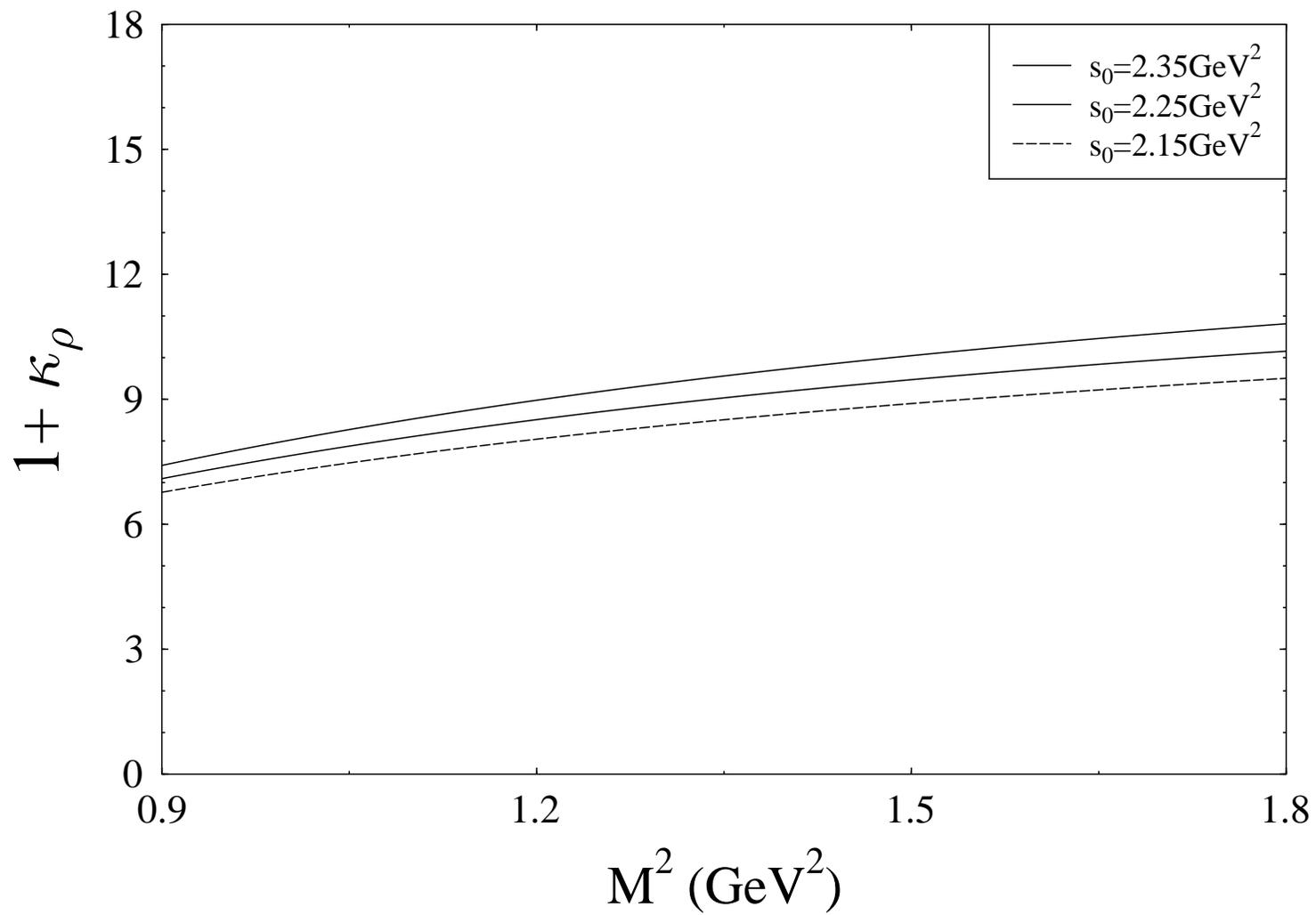
\end{document}